\documentclass{PoS}

\usepackage{epsfig,times,color,cite,rotating,picture,calc,tabularx,multirow,comment,wrapfig}
\usepackage[font=small,belowskip=-5pt,aboveskip=5pt]{caption}
\usepackage[para]{footmisc}
\usepackage{graphicx,adjustbox}
\usepackage[numbers]{natbib}
\usepackage{amsmath}
\usepackage[percent]{overpic}

\def \1{2FGL J0545.6+6018}
\def \2{2FGL J1115.0$-$0701}
\def \deg{$^{\circ}$}
\def \fermi{{\em Fermi}}

\title{Hunting for dark matter subhalos among the Fermi-LAT sources with VERITAS}

\ShortTitle{Hunting for dark matter subhalos among the Fermi-LAT sources with VERITAS}
\author{\speaker{Daniel Nieto} for the VERITAS Collaboration\thanks{http://veritas.sao.arizona.edu/} \\
        Columbia University - Nevis Laboratories\\
        E-mail: \email{nieto@nevis.columbia.edu}}

\abstract{The distribution of dark matter in the Galaxy, according to
state-of-the-art simulations, shows not only a smooth halo component
but also a rich substructure where a hierarchy of dark matter subhalos
of different masses is found. We present a search for potential dark
matter subhalos in our Galaxy exploiting the high (HE, 100 MeV -- 100
GeV) and very-high-energy (VHE, >100 GeV) $\gamma$-ray bands. We assume a
scenario where the dark matter is composed of weakly interacting
massive particles of mass over 100 GeV, and is capable of
self-annihilation into standard model products. Under such a
hypothesis, most of the photons created by the annihilation of dark
matter particles are predicted to lay in the HE $\gamma$-ray band, where
the Fermi-Large Area Telescope is the most sensitive instrument to
date. However, the distinctive spectral cut-off located at the dark
matter particle mass is expected in the VHE $\gamma$-ray band, thus
making imaging atmospheric Cherenkov telescopes like VERITAS the best
suited instruments for follow-up observations and the characterization
of a potential dark matter signature. We report on the ongoing VERITAS
program to hunt for these dark matter subhalos, particularly focusing
on two promising dark matter subhalo candidates selected among the
Fermi-LAT Second Source Catalog unassociated high-energy $\gamma$-ray
sources.}

\FullConference{The 34th International Cosmic Ray Conference,\\
		30 July- 6 August, 2015\\
		The Hague, The Netherlands}

\begin{document}
\setcitestyle{square}
\setlength\abovedisplayskip{-15pt}
\setlength\belowdisplayskip{5pt}
\setlength\abovedisplayshortskip{-15pt}
\setlength\belowdisplayshortskip{5pt}

\section{Introduction}

There are strong observational evidences that support the existence of
dark matter (DM) in our universe, like the dynamics of galaxies and
galaxy clusters, and the gravitational lensing effect. The abundance
of this unknown component of the universe has been quantified to 84\%
of its total mass-density~\cite{2015arXiv150201589P}.  Assuming the
hypothesis of DM being composed of weakly interacting massive
particles (WIMPs), annihilation or decay of DM into Standard Model
particles opens the possibility of indirectly detecting DM through the
observation of these products, amongst them $\gamma$-ray photons
(see~\cite{2015arXiv150306348C}). The expected flux of prompt
$\gamma$-ray emission from DM annihilation features a distinctive
spectral shape primarily characterized by a cut-off at the WIMP
mass. The preferred targets to search this DM spectral signature from
are those regions showing the highest signal strength, by holding a
large concentration of DM and/or being located close to the observer,
like the Galactic Center and Halo, the dwarf spheroidal satellites of
the Milky Way, and galaxy clusters. No clear DM signal has been
detected so far in any of those objects~\cite{2015arXiv150306348C}.

High-resolution simulations of Milky Way-like DM halos indicate the
presence of substructure down to the smallest resolved scales~(see,
e.g., \cite{Diemand:2008a}). Not all of these subhalos may have
accumulated significant amounts of baryonic matter and would therefore
be invisible to astronomical observations from radio to X-ray
energies. Assuming a self-annihilating WIMP in the GeV-TeV mass range,
these subhalos would be only visible at $\gamma$-ray energies. Since
the emission from WIMP annihilation is expected to be steady, such
hypothetical sources would be found in deep sky $\gamma$-ray surveys,
and could already be part of the collection of \fermi-Large Area
Telescope (LAT) sources showing no conventional counterpart at any
other wavelengths~\cite{2011PhRvD..83b3518P}. If the distinct spectral
cut-off at the WIMP mass is located at energies too high to be
measurable by \fermi-LAT, imaging atmospheric Cherenkov telescopes
(IACTs) like VERITAS may be crucial to identify such a DM
signature. Several searches for DM subhalos in the HE $\gamma$-ray
band have been conducted (see, e.g.,
~\cite{Nieto:2011c,Zechlin:2012b,2015arXiv150402087B}) and some of
them have already triggered follow-up observations by
IACTs~\cite{Nieto:2011e,2013arXiv1303.1406G}.

This contribution focuses on the observations of two DM subhalo
candidates in the HE and VHE $\gamma$-ray
bands. Section~\ref{sec:selection} describes the source selection,
while the VERITAS VHE $\gamma$-ray and \fermi-LAT HE $\gamma$-ray
observations and data analyses are contained in
Sections~\ref{sec:veritas} and~\ref{sec:fermi} respectively. In
Section~\ref{sec:results} the results are discussed. A brief summary
and outlook can be found in Section~\ref{sec:summary}.

\section{Source selection}
\label{sec:selection}

The Second \fermi-LAT Catalog (2FGL) contains 1873 HE $\gamma$-ray
sources detected by the LAT instrument after the first 24 months of
observations, 576 sources lacking any clear association. We adapted
the selection criteria from \cite{Nieto:2011e} to find good DM subhalo
candidates among the latter unassociated sources. We requested our
candidates: I. to be located at high Galactic latitudes
($|b|>10^{\circ}$); II. not to present a variable flux; III. not to
have potential counterparts at other wavelengths; IV. to be observable
from VERITAS latitude with a maximum culmination zenith angle of
40$^{\circ}$. To satisfy criterium III. we examined the 95\%
containment region of the candidates, looking for cataloged sources in
the HEASARC Archive\footnote{http://heasarc.gsfc.nasa.gov/} and
analyzed {\em Swift}-XRT\footnote{http://swift.gsfc.nasa.gov/}
exposures when available. In addition, we estimated the required
observation time for a 5$\sigma$ detection with VERITAS, based on a
2FGL Catalog flux extrapolation to the VHE range, discarding
candidates that would require more than 50 hours of VERITAS time to be
detected. We ended up with two best candidates, namely, \1 and \2. It
is worth stressing that, by the time these sources were proposed for
VERITAS observations, the 2FGL Catalog was the most updated publicly
released catalog of \fermi-LAT point like sources. Thus, 2FGL Catalog
denominations for our DM subhalo candidates will be used throughout
the text.

\section{VERITAS observations and data analysis}
\label{sec:veritas}

The Very Energetic Radiation Imaging Telescope Array
System~\cite{2006APh....25..391H} is a ground-based $\gamma$-ray
telescope array located at the Fred Lawrence Whipple Observatory in
southern Arizona (31 40N, 110 57W, 1.3 km a.s.l.). The array consists
of four imaging atmospheric-Cherenkov telescopes, each employing a
tessellated 12 m Davies-Cotton reflector instrumented with a
photomultiplier-tube camera with a 3.5\deg~field of view. VERITAS is
designed to detect emission from astrophysical objects in the energy
range from 85 GeV to greater than 30 TeV, with a nominal sensitivity
sufficient to detect, at the $5\sigma$ level, a steady point-like
source with 1$\%$ of the Crab Nebula flux in approximately 25
hrs. VERITAS has an energy resolution of 15$\%$ and an angular
resolution (68$\%$ containment) of 0.1\deg~at 1 TeV.

\begin{figure}[!t]
  \centering   
  \begin{overpic}[angle=-90,width=0.45\linewidth,clip=true,trim= 40 0 0 0]{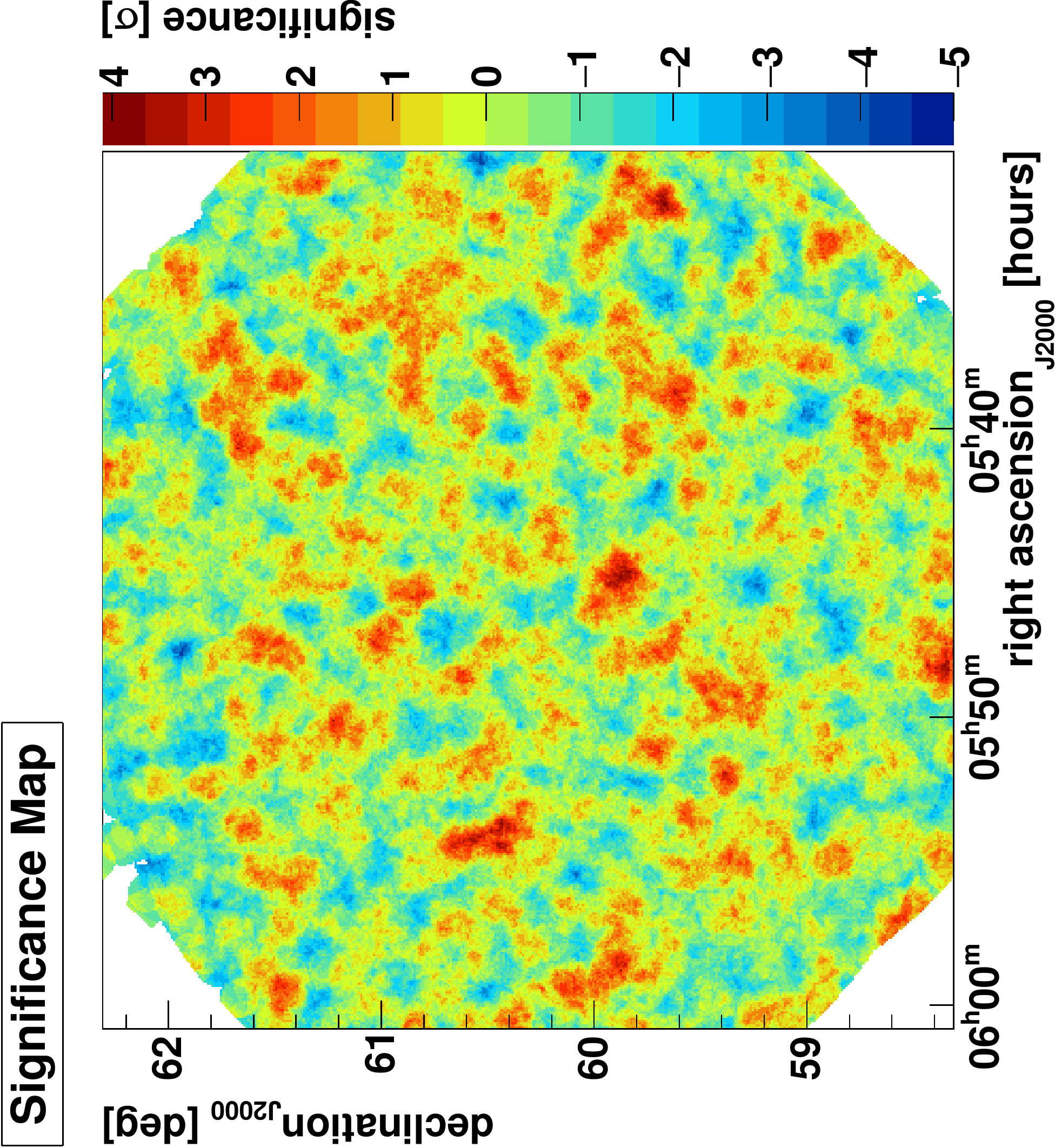}
     \put(58,10){\includegraphics[scale=0.1,clip=true,trim= 0 0 40 40]{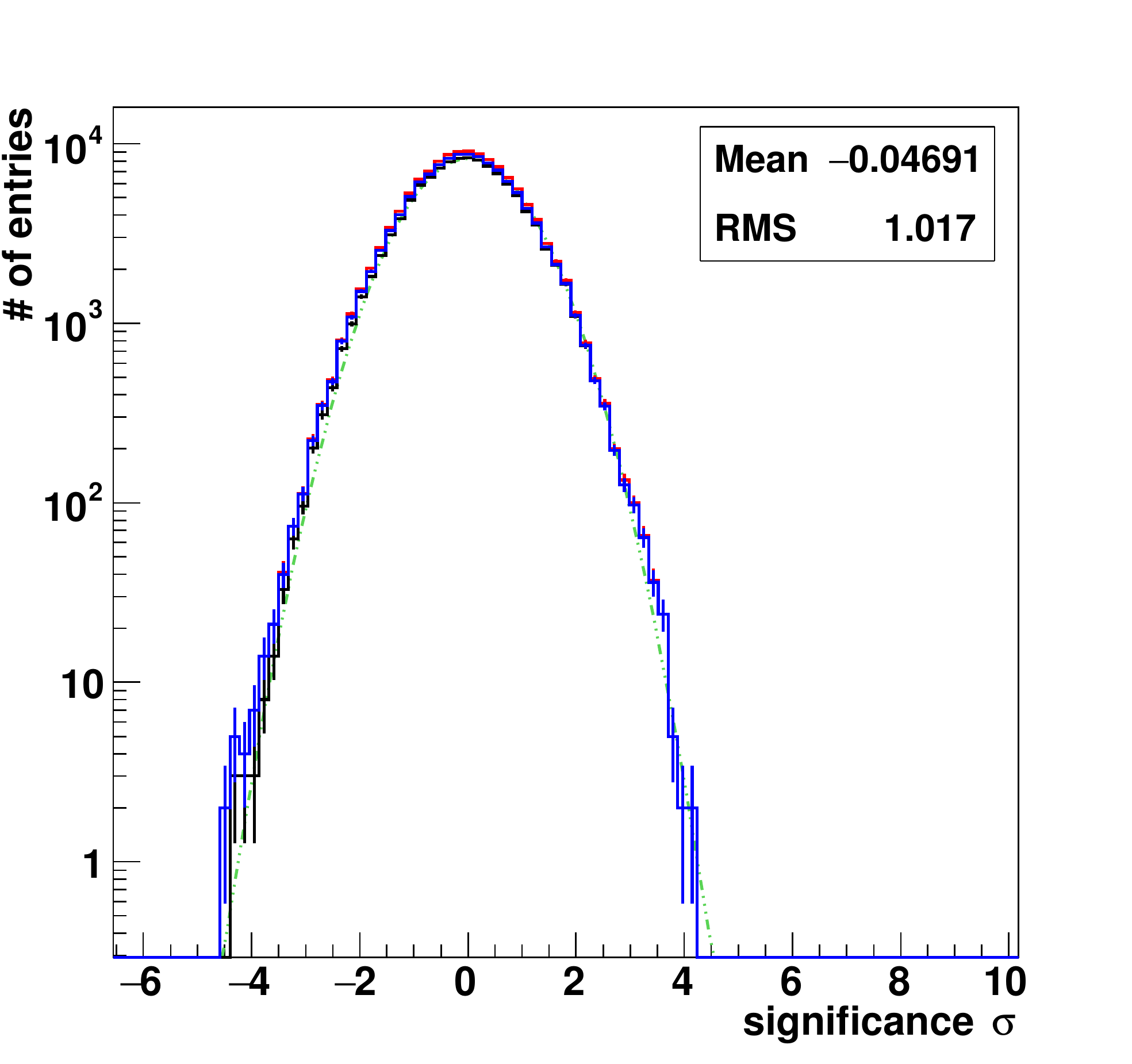}}  
     \put (10,85) {\tiny{VERITAS - ICRC 2015}}
  \end{overpic}
  \begin{overpic}[angle=-90,width=0.45\linewidth,clip=true,trim= 40 0 0 0]{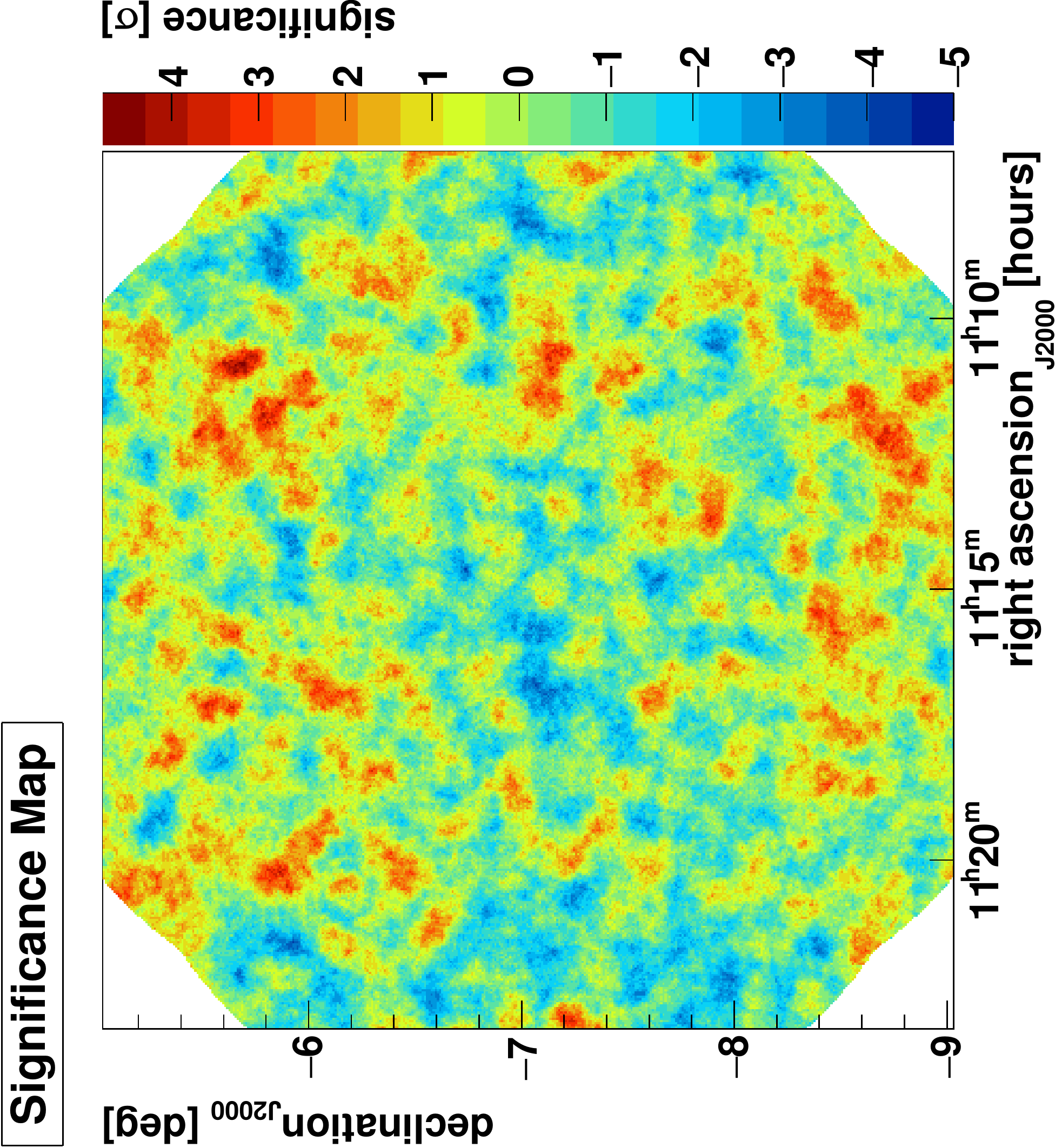}
     \put(58,10){\includegraphics[scale=0.1,clip=true,trim= 0 0 40 40]{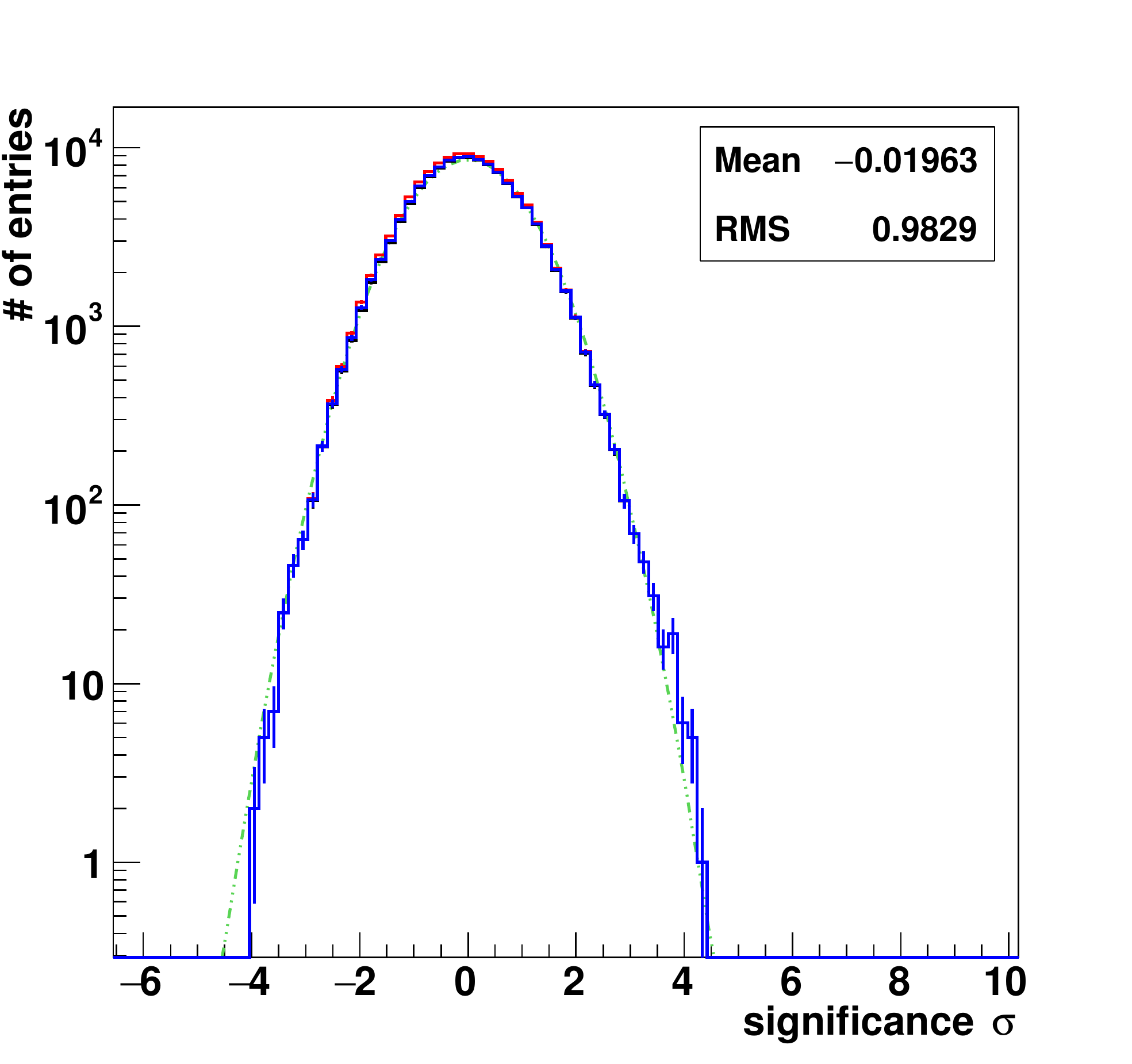}}  
     \put (10,85) {\tiny{VERITAS - ICRC 2015}}
  \end{overpic}
\caption{VERITAS significance sky maps of the regions centered on 2FGL J0545.6+6018 (left panel) and 2FGL J1115.0-0701 (right panel). Both significance distributions are well described by the hypothesis of background fluctuations, as illustrated by the significance histograms insets.}
\label{fig:vts_skymaps}
\end{figure}

The VERITAS observations of \1 and \2 presented here were made under
clear, moonless skies between October 2013 and March 2015, all of them
after the 2012 upgrade of the VERITAS cameras and trigger
system. After data-quality selection, \1 observations amount to 8.5
hrs performed at an average elevation of 61\deg, while \2 observations
span 13.8 hrs at an average elevation of 51\deg. The data were
acquired in the so called {\em wobble} (also known as {\em false
  source}) observation mode. The shower images from each telescope are
calibrated, cleaned, and parametrized using a momentum analysis. These
parameters are utilized to discriminate $\gamma$-ray like events
against hadronic events applying multivariate analysis techniques, and
are also combined to estimate the arrival direction and energy of the
original $\gamma$-ray. The event selection cuts are optimized {\em a
  priori} over a Crab Nebula data set.

We find no evidence for $\gamma$-ray emission above VERITAS energy
threshold in any of the observed fields, as illustrated by the
significance maps shown in Fig.~\ref{fig:vts_skymaps}. We note that
the distribution of significances in either fields are compatible with
pure background fluctuations.  It is worth mentioning that the
sources' coordinates in the 3FGL Catalog do not significantly change
with respect to the values provided in the 2FGL Catalog (and used for
VERITAS observations). The angular distance between both sets of
coordinates are 3 arcmin and 4 arcsec for \1 and \2 respectively, well
below VERITAS angular resolution.

The results from the VERITAS data analyses are summarized in
Table~\ref{tab:results_vts_evt}, where observations significances and
upper limits to the number of excess events coming from each source
can be found, and Table~\ref{tab:results_vts_flux}, where integral and
differential upper limits to the VHE $\gamma$-ray flux coming are
listed. For the calculation of the upper limits to the number of
excess events we applied the Rolke et
al. method~\cite{Rolke:2005a}. For the calculation of the upper limits
to the flux we assumed power-law spectra with spectral indices of -2.0
and -2.1 for \1 and \2 respectively\footnote{These spectral indices
  are the best fit values to power law spectra from dedicated LAT
  analyses presented in Sec.~\ref{sec:fermi}.}. We constrain the
integral flux of the sources in the VHE $\gamma$-ray domain to be
below 1$\%$ of the Crab Nebula integral flux at a 95$\%$ confidence
level.

\begin{table}[t]
  \centering
  \begin{adjustbox}{max width=0.8\textwidth}
    \begin{tabular}{@{} lccccccc @{}}
      \hline
      Source& $E_{th}$ [GeV]& $N_{on}$ & $N_{off}$&$N_{exc}$ & $\sigma$ & $N^{UL (95\% ~c.l.)}_{exc}$ & $N^{UL (99\%~c.l.)}_{exc}$ \\
      \hline
      \multirow{3}{*}{2FGL J0545.6+6018}&150&327 &5836 &-16.2 $\pm$ 18.6&-0.9&24.0&44.1\\
      &250&94 &	1704 &	-6.7$\pm$10.0& 	-0.7 &	28.4 &	52.0 \\
      &450&24 	&536 	&-8.1$\pm$5.1 	&-1.5 	&3.3 	&8.4 \\
      \hline
      \multirow{3}{*}{2FGL J1115.$\pm$0-0701}&150&673 	&10024 &	-4.7 $\pm$ 26.8 &	-0.2 &	99.7 	&59.2 \\
      &250&144 &	2112 &	0.5 $\pm$ 12.4 &	0.0 	&53.6 &	32.0 \\
      &450&31& 	488 	&-2.6 $\pm$ 5.8& 	-0.4 &	21.0 &	12.2 \\
      \hline
    \end{tabular}
  \end{adjustbox}
  \caption{Upper limit to the number of excess events for three
    different energy thresholds $E_{th}$. $N_{on}$ and $N_{off}$ are
    the number of gamma-like events in the signal and background
    regions respectively, $N_{exc}$ being the number of excess event
    in the signal region. The significance $\sigma$ is computed
    applying Eq. 17 in~\cite{Li:1983a}.}\label{tab:results_vts_evt}
\end{table}

\begin{table}[t]
  \centering
  \begin{adjustbox}{max width=\textwidth}
    \begin{tabular}{lcccccccc}
      \hline
      &&  \multicolumn{6}{c}{Integral} & Differential \\
      &&  \multicolumn{2}{c}{E$>$150 GeV}&\multicolumn{2}{c}{E$>$250 GeV}&\multicolumn{2}{c}{E$>$450 GeV}& E$_0$ = 1 TeV\\
      \hline
      &c.l.&  [cm$^{-2}$s$^{-1}$] & [C.U.] & [cm$^{-2}$s$^{-1}$] & [C.U.] & [cm$^{-2}$s$^{-1}$] & [C.U.] & [TeV$^{-1}$cm$^{-2}$s$^{-1}$]\\
      \hline
      \multirow{2}{*}{2FGL J0545.6+6018}&95\%&1.95$\times10^{-12}$  & 0.6\%& 0.95$\times10^{-12}$  	&0.6\%&	0.16$\times10^{-12}$ &	0.2\% &1.88$\times10^{-13}$\\
      &99\%&3.57$\times10^{-12}$  	&1.0\%	&1.69$\times10^{-12}$ 	&1.0\% 	&0.42$\times10^{-12}$ 	&0.6\%& 	2.38$\times10^{-13}$ \\
      \hline
      \multirow{2}{*}{2FGL J1115.0-0701}&95\%&2.15$\times10^{-12}$& 	0.6\% & 	0.88$\times10^{-12}$ 	&0.5\%& 	0.31$\times10^{-12}$ 	&0.4\% & 	2.11$\times10^{-13}$ \\
      &99\%&3.62$\times10^{-12}$ 	&1.0\% & 	1.48$\times10^{-12}$ &	0.9\% & 	0.53$\times10^{-12}$ &	0.8\% & 	3.54$\times10^{-13}$ \\
      \hline
    \end{tabular}
  \end{adjustbox}
  \caption{Upper limits to the integral fluxes for three different
    energy thresholds. The upper limits to the differential flux are
    computed for a pivot energy of 1
    TeV.}\label{tab:results_vts_flux}
\end{table}

\section{{\em Fermi}-LAT observations and data analysis}
\label{sec:fermi}

The LAT is a space-based electron-positron pair-conversion instrument
on board the {\em Fermi} $\gamma$-ray Space
Telescope~\cite{Atwood:2009a}. It is sensitive to $\gamma$ rays from
20 MeV to more than 300 GeV. The LAT has a field of view of 2.4 sr and
operates primarily in an all-sky survey mode, covering the entire sky
approximately every three hours.

We examined 6.7 years of LAT data (2008-07-31 to 2015-04-02, dubbed
7-years analysis henceforth) selecting reprocessed Pass 7 events in
the energy range between 100 MeV and 300 GeV. We selected class 2
events recorded at zenith angles below 100\deg~during good time
intervals. We analyzed regions of interest of 10\deg~radius centered
on the updated 3FGL coordinates of the sources. The data were analyzed
using the {\em Fermi} Science Tools version
\texttt{v9r33p0-fssc-20140520}.

\begin{figure*}[!t]
	\centering	
	\begin{overpic}[width=0.49\linewidth,clip=true,trim= 30 0 50 20]{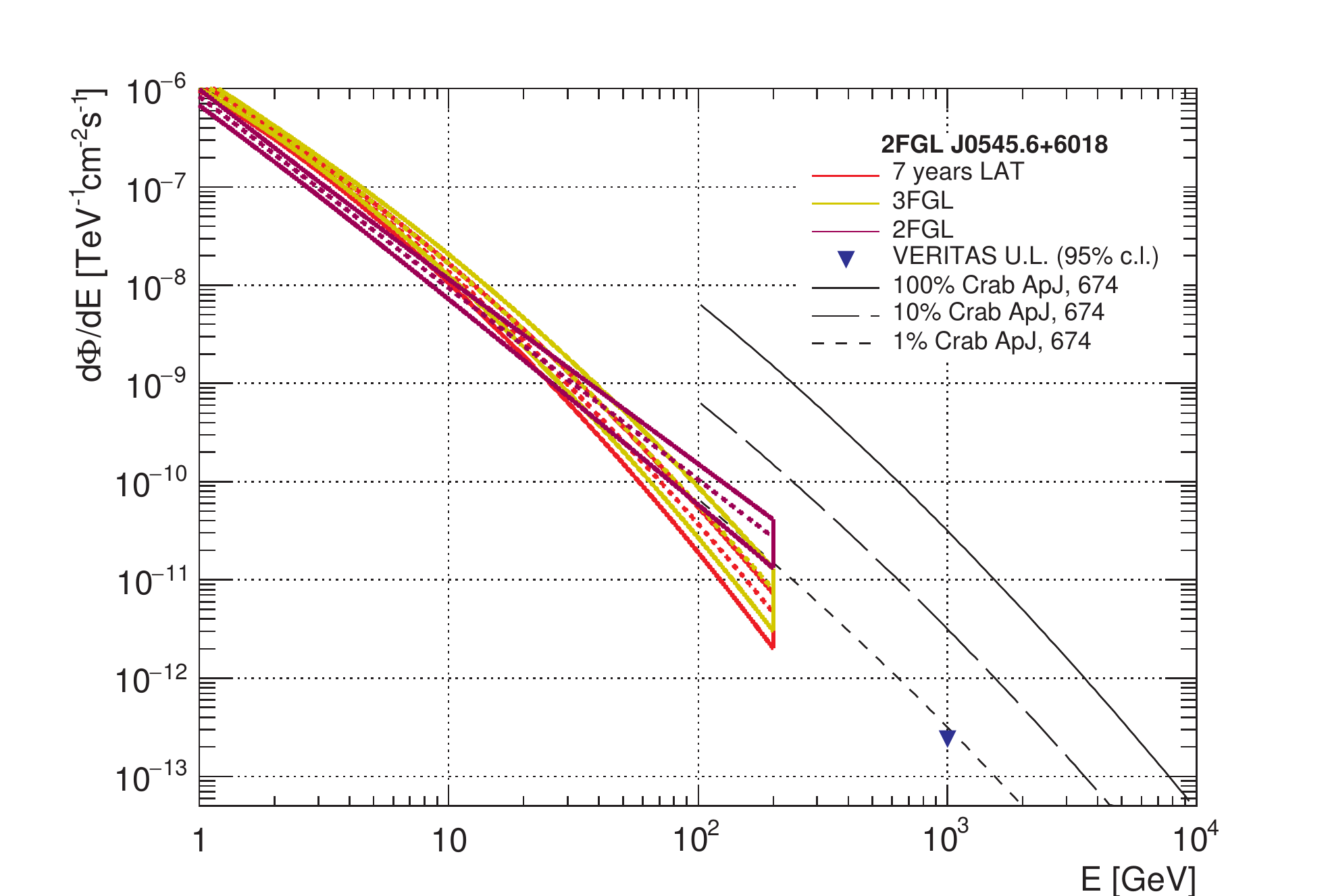}\put (11,73) {\tiny{VERITAS - ICRC 2015}}
	\end{overpic}
	\begin{overpic}[width=0.49\linewidth,clip=true,trim= 30 0 50 20]{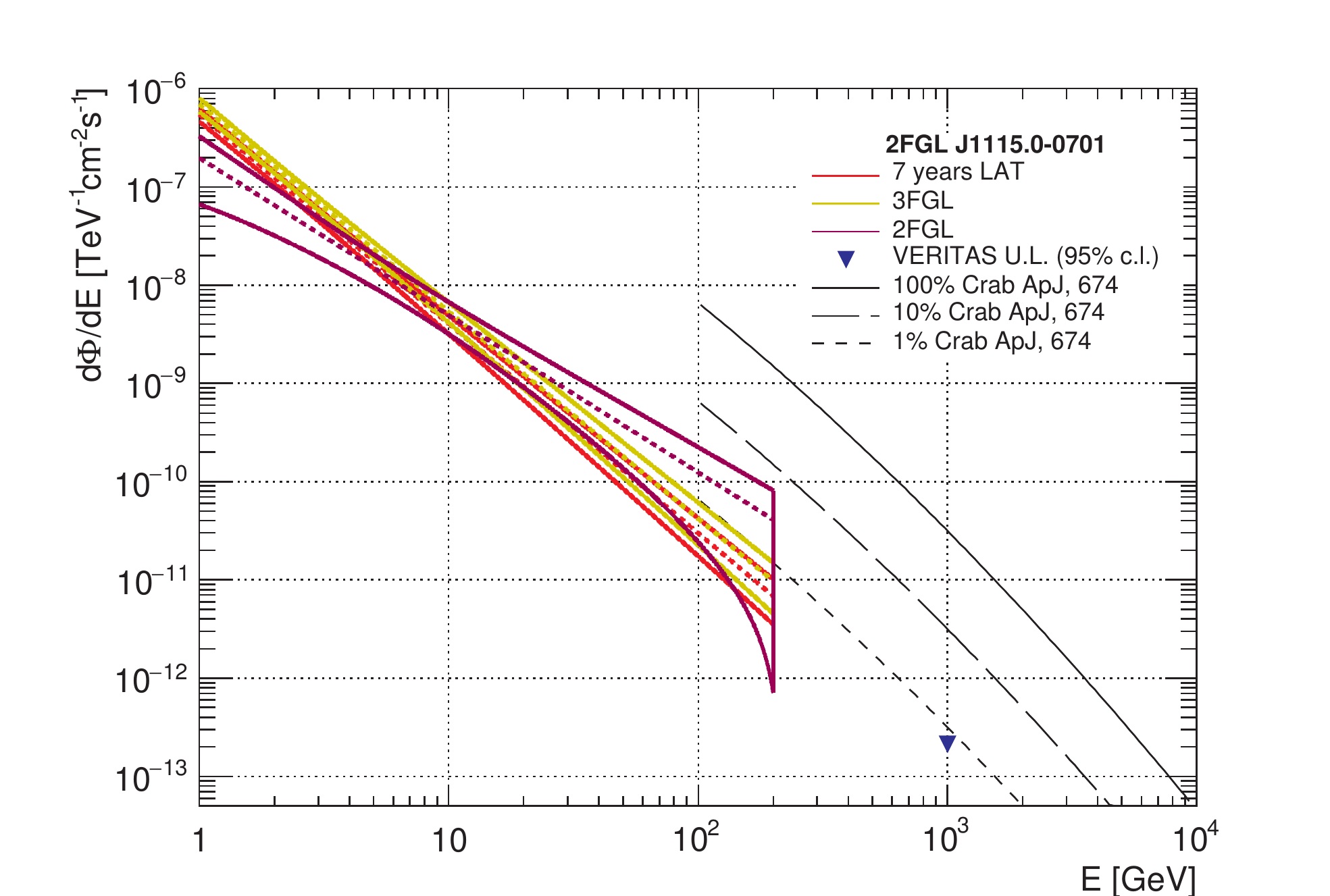}
	    	\put (11,73) {\tiny{VERITAS - ICRC 2015}}
	\end{overpic}	
	\caption{\fermi-LAT spectrum and 1$\sigma$ confidence contour
          of the fit, together with the VERITAS differential flux
          upper limits (95\% {\em c.l.}) for \1 ({\em left}) and \2
          ({\em right}). The Crab Nebula differential spectrum in the
          VHE $\gamma$-ray band~\cite{Albert:2007a} at 100\%, 10\%,
          and 1\% levels is shown as reference.}
	\label{fig:lat_spectra}
\end{figure*}

Our 7-year analysis confirms the spectral characterization found in
the 3FGL Catalog for the two sources of interest, and disfavors the
spectral characterization found in the 2FGL Catalog. The collection of
spectral parameters for both \1 and \2 are shown in
Table~\ref{tab:lat_results}. The agreement between our analyses and
the 3FGL Catalog is also illustrated in Fig.~\ref{fig:lat_spectra}.

In addition, we independently confirm the variable nature of the
$\gamma$-ray emission from \2 as first reported in the 3FGL
Catalog. More specifically, the source presented a high activity state
in mid January 2011 lasting for approximately two weeks with integral
fluxes of $\sim4\times10^{-9}$ cm$^{-2}$ s$^{-1}$ above 1 GeV
($TS=37$), while its $TS$ value for the last 12 months of analyzed
data is below 10 for energies larger than 100 MeV. The difference
(>1$\sigma$) in the normalization factors between our analysis and the
3FGL Catalog nominal value could be attributed to the apparent fade
off of \2, while the spectral indices are compatible within errors.

\begin{table}[t]
  \centering
  \begin{adjustbox}{max width=0.8\textwidth}
    \begin{tabular}{lccccc}
      \hline
      &Analysis&$E_0$&$N$&$\alpha$&$\beta$\\
      &&[MeV]&[$\times10^{-13}$ MeV$^{-1}$ cm$^{-2}$ s$^{-1}$] & &\\
      \hline
      \multirow{3}{*}{\1}&	2FGL &2292&(1.6 $\pm$ 0.3)$\times10^{-13}$& 1.95 $\pm$ 0.11 & n/a\\      
      &3FGL & 1913&(4.0 $\pm$ 0.5)$\times10^{-13}$&1.79 $\pm$ 0.13&0.30 $\pm$ 0.09\\
      &7-Years & 1913 & (3.7 $\pm$ 0.4)$\times10^{-13}$&1.76 $\pm$ 0.12&0.33 $\pm$ 0.08 \\
      \hline
      \multirow{3}{*}{\2}&	2FGL &7491&(7.8 $\pm$ 2.7)$\times10^{-15}$&1.60 $\pm$ 0.28& n/a\\
      &3FGL & 1947&(1.7 $\pm$ 0.2)$\times10^{-13}$ & 2.11 $\pm$ 0.11 & n/a\\
      &7-Years & 1947 & (1.3 $\pm$ 0.1)$\times10^{-13}$ &2.13 $\pm$ 0.10& n/a\\
      \hline
    \end{tabular}
  \end{adjustbox}
  \caption{Spectral parameters for \1 and \2 coming from the 2FGL and
    3FGL Catalogs as well as our dedicated 7-year analysis. For
    sources showing a power-law best fit to their spectra these are
    parametrized as $dN/dE=N(E/E_0)^{-\alpha}$, while for sources
    best described by a log-parabola their spectra are parametrized
    as $dN/dE=N(E/E_0)^{-\alpha-\beta log(E/E_0)}$. All errors are
    statistical.}\label{tab:lat_results}
\end{table}

\section{Results}
\label{sec:results}

The flux variability for \2 first noted in the 3FGL Catalog and later
verified by our dedicated LAT analysis completely rejects the
hypothesis of this source being a DM subhalo. Our 7-years analysis
showed a high flux state with approximately two weeks duration
compatible with an integral flux above 1 GeV of $\sim4\times10^{-9}$
cm$^{-2}$ s$^{-1}$ (14\% of the inverse Compton component of the Crab
Nebula in the HE $\gamma$-ray band). VERITAS constraints to \2 VHE
flux would rule out the strict extrapolation of the 2FGL and 3FGL
spectra into the VHE domain as described in
Table~\ref{tab:lat_results}. The extrapolation of the 7-years spectral
characterization is not ruled out by the VERITAS constraints but it is
in strong tension with them. In addition, VERITAS upper limits are
still compatible within the extrapolated 1$\sigma$ confidence contour
coming from each of those analyses. The tension between the VERITAS
limits and our 7-years LAT analysis might favor some spectral
softening at VHE which, if attributed to attenuation due to
extragalactic background light absorption, together with the variable
HE flux of the source and its high Galactic latitude, might support a
distant blazar hypothesis as the nature of this source. However, it is
also possible that the source was actually in a low state during part
or all of the VERITAS observations, correlated to the lack of activity
observed during the last 12 months of analyzed LAT data. In such a
scenario nothing could be said in regards to the spectral softening of
\2 in the VHE band.

While VERITAS limits rule out the direct extrapolation to the VHE
domain of the 2FGL Catalog spectral description of \1, they do not
constrain at all the log parabola description from the 3FGL Catalog
and our 7-years analysis. Following up with the hypothesis of \1 being
a DM subhalo we interpret its SED in the $\gamma$-ray band to be
purely originated by annihilation of WIMPs in the object. We model the
differential flux as in Eq.~\ref{eq:dm_diff}, where
$\langle\sigma_{ann} v\rangle$ is the thermally-averaged
self-annihilation cross-section, $m_{WIMP}$ is the WIMP mass,
$dN^i_\gamma/dE$ is the photon yield per annihilation through channel
$i$, $B^i$ is the branching ratio for channel $i$, and $\rho$
represents the density distribution of DM. We assume a value for the
thermally-averaged self-annihilation cross-section of
$\langle\sigma_{ann} v\rangle=2.2\times10^{-26}$ cm$^3$
s$^{-1}$~\cite{2012PhRvD..86b3506S} and consider the parametrizations
from~\cite{Cembranos:2010a} as coded in the DAMASCO
package\footnote{http://cta.gae.ucm.es/gae/damasco} for the photon
yield per annihilation. Thus, our model contains two free parameters,
$m_{WIMP}$ from the {\em particle physics factor} and the {\em
  astrophysical factor} $J$.

\begin{equation}
\phi(E,\Delta\Omega) = \phi^{PP}(E) \times J(\Delta\Omega)=
\underbrace{\frac{1}{4\pi}\frac{\langle\sigma_{ann}v\rangle}{2
    m^2_{DM}}\sum^n_{i=1} B^i
  \frac{dN^i_\gamma}{dE}dE}_\text{Particle physics factor} \times
\underbrace{\int_{\Delta\Omega,~l.o.s.}\rho^2(r(s,\Omega)) ds
  d\Omega}_{\text{Astrophysical factor}}.
\label{eq:dm_diff}
\end{equation}

\begin{wraptable}{r}{0.5\textwidth}
  \centering
  \begin{adjustbox}{max width=0.5\textwidth}
    \begin{tabular}{@{} lcccc @{}}       \hline
      Channel & $m_{WIMP}$& log$_{10}(J)$ & $\chi^{2}/d.o.f.$ & $p-value$\\
      & [GeV] &  [log$_{10}$(GeV$^2$ cm$^{-5}$)] & & \\
      \hline
      $b\bar{b}$& 78.3 $\pm$ 11.6& $20.8^{+0.2}_{-0.3}$& 1.02 & 0.36\\
      $W^{+}W^{-}$&89.2 $\pm$ 13.7 & $20.9^{+0.2}_{-0.3}$& 0.74 & 0.48 \\
      $\tau^{+}\tau^{-}$& 18.0 $\pm$ 0.3 & $20.3^{+0.1}_{-0.1}$ & 14.69 & $4\times10^{-7}$\\
      \hline
    \end{tabular}
  \end{adjustbox}
  \caption{Summary of \1 SED fit to DM annihilation spectra. We
    assume $\langle\sigma_{ann} v\rangle=2.2\times10^{-26}$ cm$^3$
    s$^{-1}$ for the estimation of the
    $J$-factors.}\label{tab:results_dm}
\end{wraptable}

Four channels are considered: annihilation into
$b\bar{b}$,~$W^{+}W^{-}$,~$\mu^+\mu^-$, and~$\tau^+\tau^-$, with
branching ratios $B^i=1$ in all scenarios. While the fit did not
converge for the $\mu^+\mu^-$ channel and the goodness of fit for the
~$\tau^+\tau^-$ is really poor (the hypothesis of DM annihilating into
$\tau$ particles is rejected at $\alpha<0.01$), the fits to
annihilation into $b\bar{b}$,~$W^+W^-$ provide with decently good
fits. The best-fit values for the WIMP mass and the {\em astrophysical
  factor} together with their statistical errors can be found in
Table~\ref{tab:results_dm}.

\begin{wrapfigure}{r}{0.5\textwidth}
  \begin{center}
    \begin{overpic}[width=0.5\textwidth,clip=true,trim= 25 0 50 40]{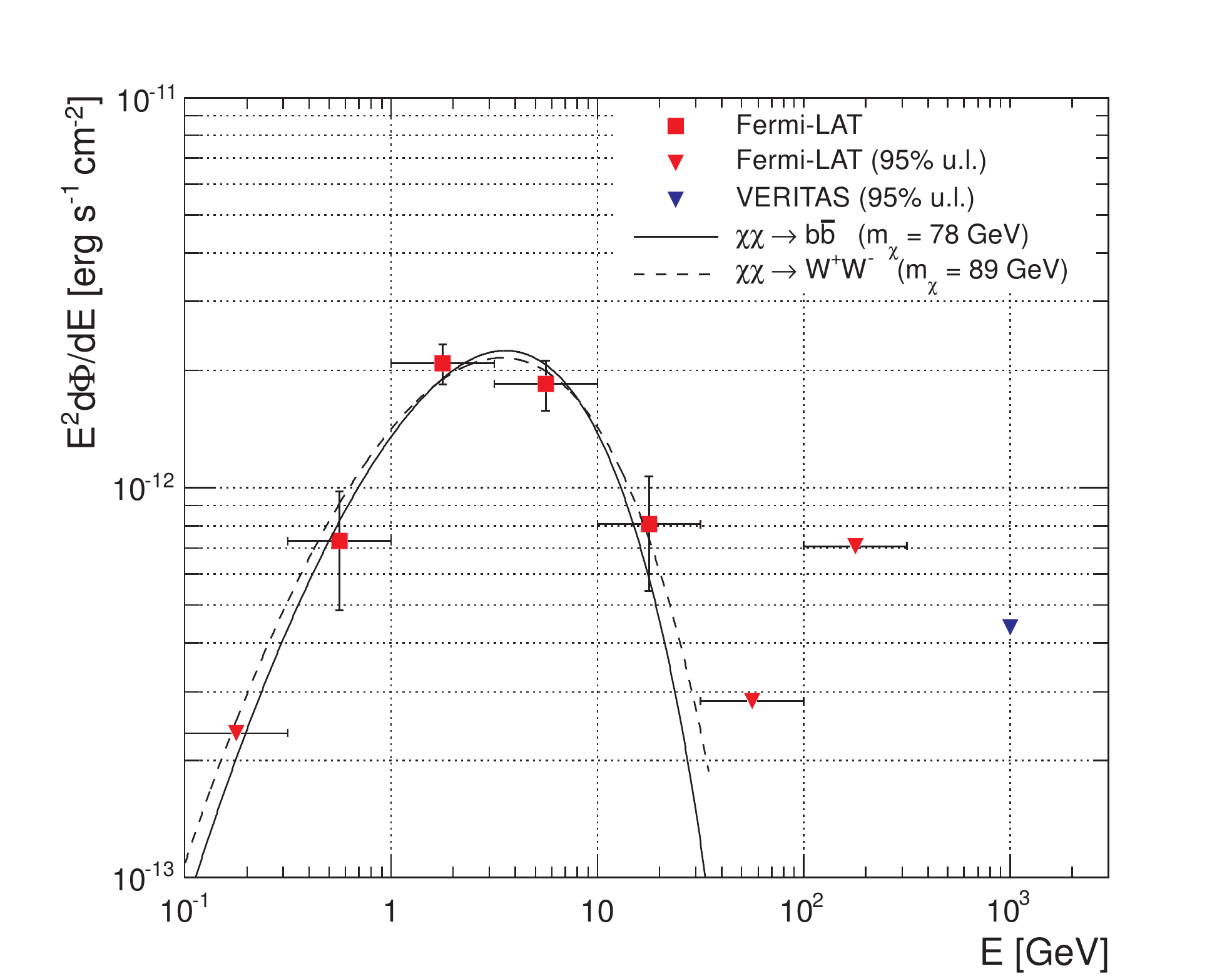}
    	\put (13,84) {\tiny{VERITAS - ICRC 2015}}
    \end{overpic}
  \end{center}
  \caption{\1 spectral energy distribution. The two best-fit
    annihilation spectra are shown.}\label{fig:dmfit}
\end{wrapfigure}

The most favored annihilation channel in terms of the goodness-of-fit
is ~$W^+W^-$, with a $\chi^{2}_{d.o.f}=0.74$, although the fit result
is in slight tension with the lowest-energy upper limit in the SED as
showed in Fig.~\ref{fig:dmfit}. On the other hand, annihilation into
$b\bar{b}$, with a $\chi^{2}_{d.o.f}=1.04$, is compatible with the
lowest-energy upper limit in the SED. We note that the later
annihilation channel and its associated WIMP mass of $\sim89$ GeV are
similar to the results presented in~\cite{2015arXiv150402087B} for a
different set of DM subhalo candidates, and in reasonable agreement
with the \fermi-LAT Galactic Center excess DM modeling
in~\cite{2015PhRvD..91f3003C}.  The estimated $J$-factors are one
order of magnitude above the nominal $J$-factor for the most promising
ultra-faint dwarf spheroidal galaxy in~\cite{Bonnivard:2015xpq},
namely Ursa Mayor II\footnote{log$_{10}(J_{\text{UMa~II}}) =
  19.9^{+0.7}_{-0.5}$ for an integration angle of 0.5\deg~ (the
  $J$-factor units are GeV$^2$ cm$^{-5}$).}.The most constraining
limits in this energy range come from the analysis of \fermi-LAT data
on dwarf spheroidal galaxies~\cite{2015arXiv150302641F}. For
annihilation into $b\bar{b}$ and $m_{WIMP}=78$~GeV, they found
$\langle\sigma_{ann} v\rangle<1.0\times10^{-25}$ cm$^3$ s$^{-1}$ at
95$\%$ {\em cl}. Similarly, for annihilation into $W^{+}W^{-}$ and
$m_{WIMP}=89$~GeV, they found $\langle\sigma_{ann}
v\rangle<1.6\times10^{-25}$ cm$^3$ s$^{-1}$ at 95$\%$ {\em
  cl}. Therefore, although in tension with the latter limits, our
results are not ruled out by them.

\section{Summary and outlook}

\label{sec:summary}

We have presented a search for DM subhalo candidates in the 2FGL
Catalog and the results from the observations with VERITAS of two
promising candidates, \1 and \2, along with a dedicated 7-years LAT
data for both sources.

\2 flux variability and high Galactic latitude opens the possibility
of this source being a blazar. We encourage the monitoring of \2 HE
$\gamma$-ray activity searching for potential flaring events which
could drive its VHE $\gamma$-ray component, depending on the source's
redshift, into the sensitivity reach of the current generation of
IACTs.

We interpreted \1 SED in terms of DM annihilating via different
channels. Annihilation into $b\bar{b}$ and ~$W^+W^-$ fit the source's
$\gamma$-ray emission reasonably well, providing WIMP masses of 78.3
GeV, and 89.2 GeV correspondingly. While \1 $\gamma$-ray emission has
been interpreted in terms of a DM signal, less exotic explanations
compatible with its spectral shape can be offered, like attributing
its emission to a radio-faint $\gamma$-ray pulsar. Multiwavelength
follow-up observations are encouraged in order to elucidate the real
nature of this enigmatic source.

It is worth mentioning the impact that the next-generation of IACTs,
represented by the Cherenkov Telescope Array
(CTA,~\cite{Acharya20133}), may have on this kind of searches. CTA is
expected to improve current-generation IACTs' sensitivity by a factor
of ten which, in conjunction with its survey capabilities, will
increase the likelihood to serendipitously discover new VHE
$\gamma$-ray sources and, potentially, DM subhalos.

\section{Acknowledgments}
{This research with VERITAS is supported by grants from the
  U.S. Department of Energy Office of Science, the U.S. National
  Science Foundation and the Smithsonian Institution, and by NSERC in
  Canada. We acknowledge the excellent work of the technical support
  staff at the Fred Lawrence Whipple Observatory and at the
  collaborating institutions in the construction and operation of the
  instrument. DN acknowledges partial support from the National
  Aeronautics and Space Administration through the {\em Fermi} Guest
  Investigator Program. The VERITAS Collaboration is grateful to
  Trevor Weekes for his seminal contributions and leadership in the
  field of VHE gamma-ray astrophysics, which made this study
  possible.}

{\scriptsize 
\setlength{\bibsep}{0pt}
\bibliography{references}{}

\providecommand{\href}[2]{#2}\begingroup\raggedright\begin{thebibliography}{10}

\bibitem{2015arXiv150201589P}
{Planck Collaboration}, P.~A.~R. {Ade}, et~al., {\it {Planck 2015 results.
  XIII. Cosmological parameters}},  {\em ArXiv e-prints} (Feb., 2015)
  [\href{http://arxiv.org/abs/1502.01589}{{\tt arXiv:1502.01589}}].

\bibitem{2015arXiv150306348C}
J.~{Conrad}, J.~{Cohen-Tanugi}, and L.~E. {Strigari}, {\it {WIMP searches with
  gamma rays in the Fermi era: challenges, methods and results}},  {\em ArXiv
  e-prints} (Mar., 2015) [\href{http://arxiv.org/abs/1503.06348}{{\tt
  arXiv:1503.06348}}].

\bibitem{Diemand:2008a}
J.~Diemand, M.~Kuhlen, et~al., {\it {Clumps and streams in the local dark
  matter distribution}},  {\em Nature} {\bf 454} (2008) 735--738,
  [\href{http://arxiv.org/abs/0805.1244}{{\tt arXiv:0805.1244}}].

\bibitem{2011PhRvD..83b3518P}
L.~{Pieri}, J.~{Lavalle}, G.~{Bertone}, and E.~{Branchini}, {\it {Implications
  of high-resolution simulations on indirect dark matter searches}},  {\em
  Phys.Rev.D} {\bf 83} (Jan., 2011) 023518,
  [\href{http://arxiv.org/abs/0908.0195}{{\tt arXiv:0908.0195}}].

\bibitem{Nieto:2011c}
D.~{Nieto}, V.~{Mart{\'{\i}}nez}, et~al., {\it {A search for possible dark
  matter subhalos as IACT targets in the First Fermi-LAT Source Catalog}},
  {\em $3^{rd}$ Fermi Symposium, Rome} (Oct., 2011)
  [\href{http://arxiv.org/abs/1110.4744}{{\tt arXiv:1110.4744}}].

\bibitem{Zechlin:2012b}
H.-S. {Zechlin} and D.~{Horns}, {\it {Unidentified sources in the Fermi-LAT
  second source catalog: the case for DM subhalos}},  {\em JCAP} {\bf 11}
  (Nov., 2012) 50, [\href{http://arxiv.org/abs/1210.3852}{{\tt
  arXiv:1210.3852}}].

\bibitem{2015arXiv150402087B}
B.~{Bertoni}, D.~{Hooper}, and T.~{Linden}, {\it {Examining The Fermi-LAT Third
  Source Catalog In Search Of Dark Matter Subhalos}},  {\em ArXiv e-prints}
  (Apr., 2015) [\href{http://arxiv.org/abs/1504.02087}{{\tt arXiv:1504.02087}}].

\bibitem{Nieto:2011e}
D.~{Nieto} et~al., {\it {The search for galactic dark matter clump candidates
  with Fermi and MAGIC}},  {\em {32$^{nd}$ International Cosmic Ray Conference,
  Beijing}} (Aug., 2011) [\href{http://arxiv.org/abs/1109.5935}{{\tt
  arXiv:1109.5935}}].

\bibitem{2013arXiv1303.1406G}
A.~{Geringer-Sameth} and {for the VERITAS Collaboration}, {\it {The VERITAS
  Dark Matter Program}},  {\em 4$^{th}$ Fermi Symposium, Monterrey, CA} (2012)
  [\href{http://arxiv.org/abs/1303.1406}{{\tt arXiv:1303.1406}}].

\bibitem{2006APh....25..391H}
J.~{Holder} et~al., {\it {The first VERITAS telescope}},  {\em Astroparticle
  Physics} {\bf 25} (July, 2006) 391--401,
  [\href{http://arxiv.org/abs/astro-ph/0604119}{{\tt astro-ph/0604119}}].

\bibitem{Rolke:2005a}
W.~A. Rolke, A.~M. Lopez, and J.~Conrad, {\it {Limits and confidence intervals
  in the presence of nuisance parameters}},  {\em Nucl.Instrum.Meth.} {\bf A551} (2005)
  493--503, [\href{http://arxiv.org/abs/physics/0403059}{{\tt
  physics/0403059}}].

\bibitem{Li:1983a}
T.-P. Li and Y.-Q. Ma, {\it {Analysis methods for results in gamma-ray
  astronomy}},  {\em ApJ} {\bf 272} (1983) 317--324.

\bibitem{Atwood:2009a}
{\bf Fermi-LAT} Collaboration, W.~Atwood et~al., {\it {The Large Area Telescope
  on the Fermi Gamma-ray Space Telescope Mission}},  {\em ApJ} {\bf 697} (2009)
  1071--1102, [\href{http://arxiv.org/abs/0902.1089}{{\tt arXiv:0902.1089}}].

\bibitem{Albert:2007a}
{\bf MAGIC} Collaboration, J.~Albert et~al., {\it {VHE Gamma-Ray Observation of
  the Crab Nebula and Pulsar with MAGIC}},  {\em ApJ} {\bf 674} (2008)
  1037--1055, [\href{http://arxiv.org/abs/0705.3244}{{\tt
  arXiv:0705.3244}}].

\bibitem{2012PhRvD..86b3506S}
G.~{Steigman}, B.~{Dasgupta}, and J.~F. {Beacom}, {\it {Precise relic WIMP
  abundance and its impact on searches for dark matter annihilation}},  {\em
  Phys. Rev. D} {\bf 86} (July, 2012) 023506,
  [\href{http://arxiv.org/abs/1204.3622}{{\tt arXiv:1204.3622}}].

\bibitem{Cembranos:2010a}
J.~A.~R. {Cembranos}, A.~{de La Cruz-Dombriz}, et~al., {\it {Photon spectra
  from WIMP annihilation}},  {\em Phys. Rev. D} {\bf 83} (Apr., 2011) 083507--+,
  [\href{http://arxiv.org/abs/1009.4936}{{\tt arXiv:1009.4936}}].

\bibitem{2015PhRvD..91f3003C}
F.~{Calore}, I.~{Cholis}, C.~{McCabe}, and C.~{Weniger}, {\it {A tale of tails:
  Dark matter interpretations of the Fermi GeV excess in light of background
  model systematics}},  {\em Phys. Rev. D} {\bf 91} (Mar., 2015) 063003,
  [\href{http://arxiv.org/abs/1411.4647}{{\tt arXiv:1411.4647}}].

\bibitem{Bonnivard:2015xpq} V.~Bonnivard, C.~Combet, et~al., {\it
  {Dark matter annihilation and decay in dwarf spheroidal galaxies:
    The classical and ultrafaint dSphs}}, {\em ArXiv e-prints} (Apr., 2015),
  [\href{http://arxiv.org/abs/1504.02048}{{\tt arXiv:1504.02048}}].

\bibitem{2015arXiv150302641F}
{Fermi-LAT Collaboration}, {\it {Searching for Dark Matter Annihilation from
  Milky Way Dwarf Spheroidal Galaxies with Six Years of Fermi-LAT Data}},  {\em
  ArXiv e-prints} (Mar., 2015) [\href{http://arxiv.org/abs/1503.02641}{{\tt
  arXiv:1503.02641}}].

\bibitem{Acharya20133}
B.~Acharya et~al., {\it Introducing the CTA concept},  {\em Astroparticle
  Physics} {\bf 43} (2013), no.~0 3 -- 18. Seeing the High-Energy Universe with
  the Cherenkov Telescope Array - The Science Explored with the CTA.

\end{thebibliography}\endgroup
\bibliographystyle{JHEPs}
}
\end{document}